\documentclass[twocolumn,floatfix]{revtex4}
\usepackage{bm,psfrag, amsmath, color}

\usepackage[dvipsnames]{xcolor}
\usepackage{graphicx}
\usepackage{subfigure}

\newcommand{\beq}{\begin{equation}}
\newcommand{\eeq}{\end{equation}}
\newcommand{\beqa}{\begin{eqnarray}}
\newcommand{\eeqa}{\end{eqnarray}}
\newcommand{\ba}{\begin{array}}
\newcommand{\ea}{\end{array}}

\usepackage{mathtools,slashed}
\usepackage{youngtab}

\newcommand{\be}{\begin{equation}}
\newcommand{\ee}{\end{equation}}
\newcommand{\bea}{\begin{eqnarray}}
\newcommand{\eea}{\end{eqnarray}}

\newcommand{\GHC}{G_{\mathrm{HC}}}
\newcommand{\GSM}{G_{\mathrm{SM}}}

\newcommand{\GCUS}{G_{\mathrm{cus.}}}
\newcommand{\GF}{G_{\mathrm{F}}}
\newcommand{\HF}{H_{\mathrm{F}}}

\renewcommand\[{\left[}
\renewcommand\]{\right]}

\newcommand{\exclude}[1]{}

\def\beq{\begin{equation}}
\def\eeq{\end{equation}}
\newcommand{\C}[1]{\mathcal{#1}}

\usepackage[english]{babel}
\usepackage[utf8x]{inputenc}
\usepackage[T1]{fontenc}

\usepackage[colorinlistoftodos]{todonotes}
\usepackage[colorlinks=true, allcolors=blue]{hyperref}

\begin{document}

\title{
Small instantons \& the strong CP problem  in composite Higgs models
}
\author{R.S.~Gupta, V.V.~Khoze and M.~Spannowsky}
\affiliation{Institute for Particle Physics Phenomenology,
Durham University, South Road, Durham, DH1 3LE}

\date{\today}

\begin{abstract}
\noindent We show that QCD instantons can generate large effects at small length scales in the ultraviolet in standard composite Higgs models that utilise partial compositeness. This has important implications for possible  solutions of the strong CP problem in these models. First we show that in the simplest known UV completions of composite Higgs models, if an axion is also present, it can have a mass much larger than the usual QCD axion. Even more remarkable is the case where there are no axions, but the strong CP problem can be solved by generating the up quark mass entirely from the contribution of instantons thus reviving the massless up-quark solution for these models. In both cases no additional field content is required apart from what is required to realise partial compositeness.
\end{abstract}

\maketitle

\section{Introduction}

The strong CP problem is one of the five major particle physics puzzles  that motivate the existence of new physics beyond the Standard Model (SM). The most elegant solution to this problem is the existence of a new $U(1)$ symmetry that is anomalous under QCD so that  it is possible to rotate away the strong CP phase. The simplest possibility is  that the up quark is massless which leads to the existence of an axial $U(1)$ symmetry that makes the strong CP phase unphysical. Another possibility is the existence of the Peccei-Quinn $U(1)$ symmetry that is spontaneously broken resulting in a Goldstone mode, the axion. As the Peccei-Quinn symmetry is anomalous under QCD, the axion gets a potential due to non-perturbative QCD effects and stabilises at a value that leads to a vanishing strong CP phase~\cite{Peccei:1977hh,Weinberg:1977ma,Wilczek:1977pj}. 

Both the above solutions are thought to have unambiguous low energy consequences.  
The massless up quark solution~\cite{Georgi:1981be,Choi:1988sy,Kaplan:1986ru,Banks:1994yg,Dine:2014dga} can be tested by lattice simulations. Unfortunately the latest lattice studies indicate a non-zero up mass, that seemingly falsifies this possibility~\cite{Aoki:2016frl,Alexandrou:2020bkd}. This leaves the Peccei-Quinn solution which predicts the existence of the axion with a  mass and coupling that is restricted to lie in a narrow band in the parameter space.   A vigorous experimental effort that aims to probe the full band is currently underway.

The above predictions, however, rely on the tacit assumption that any non-perturbative contribution to the up mass  in the first case or to the axion in the second case, arises from the large instantons in the IR. If small instantons in the UV also become important it will completely alter  the above experimental expectations. Previous attempts to enhance these UV contributions to the axion mass  require additional elements-- such as new coloured fermions~\cite{randall, peskin}, extra dimensions~\cite{Gherghetta:2020keg} or a UV modification of the QCD gauge group~\cite{agrawal1,agrawal2,Csaki:2019vte}. 

In this work we show that small instanton contributions can become important in  composite Higgs models with partially composite fermions~\cite{mchm, wulzer}. This can be achieved  with  no additional field content other than what is necessary to fully realise partial compositeness in standard UV completions of these models.

  The enhancement of small instanton effects in composite model is possible  because the two factors that suppress small instanton contributions in the SM , namely,  the smallness of the strong coupling in the UV and the smallness of the product of the SM  Yukawa couplings, can both be overcome in these models. The first suppression factor can be overcome because, as we will show, in order to generate composite partners for all SM fermions, many new coloured fermions need to be introduced. These new  degrees of freedom alter the running of the QCD strong coupling in the UV where it grows again to non-perturbative values. As far as the suppression due to the Yukawa couplings is concerned, this can be overcome because in these models the effective SM Yukawa matrices can be anarchic and ${\cal O}(1)$ in the UV. 
   
   We  show that the enhancement of the small instanton contributions can be so effective in these models that it may be possible to generate the entire mass of the up quark from instanton effects. This leads to a solution of the strong CP problem as in the deep UV the up Yukawa is absent and indeed an additional $U(1)$ symmetry related to the axial rotation of left and right handed up quarks exists; such a chiral rotation can be used to completely rotate away the strong CP phase. We also show that in an alternative scenario where an axion field exists, its mass would lie outside the usual band for the QCD axion because of the enhancement of strong instanton effects in these models.  

.
\section{Model}
\label{model}
We consider a  straightforward  extension of one of the simplest known UV completions of composite models~\cite{mchm, wulzer} by Ferretti~\cite{Ferretti:2014qta} where the confining hypercolor gauge group is $SU(4)_{HC}$. The field content of our  model is shown in Table~\ref{fields}.    The original model in Ref.~\cite{Ferretti:2014qta} only has a single pair of fermions,  ${\chi}_u$ and $\tilde{\chi}_u$, and it can generate partners only for one left handed doublet and one right handed up-type quark. In order to obtain partners for all SM fermions we have  extended the field content of the original model    by simply taking three copies,  ${\chi}^i_u$ and $\tilde{\chi}^i_u$ with $i=1$-$3$, and also introducing the analogous fields $\chi^j_d$ and $\tilde{\chi}^j_d$  that will give rise to partners for the down-type SM fermions  The   Lagrangian, 
\begin{eqnarray}\label{kinetic}
{\cal L}_k= -i (\bar{{\chi}}^i_u\slashed{D}{\chi}^i_u+ \bar{{\chi}}^j_d\slashed{D}{\chi}^j_d+ \bar{\tilde{\chi}}^i_u\slashed{D}\tilde{\chi}^i_u+ \bar{\tilde{\chi}}^j_d\slashed{D}\tilde{\chi}^j_d+ \bar{\psi}\slashed{D}\psi)
\end{eqnarray}
thus has an additional $SU(3)^u_{F}\times SU(3)^d_{F}$ symmetry not present in the model in Ref.~\cite{Ferretti:2014qta}. Note that in our notation above we have only made the index, $\{i,j\}$,  corresponding  to the   $SU(3)^u_{F}\times SU(3)^d_{F}$ symmetry explicit.
\subsection{Running of hypercolour coupling}

\begin{figure}[t]
  \centering
  \includegraphics[scale=0.4]{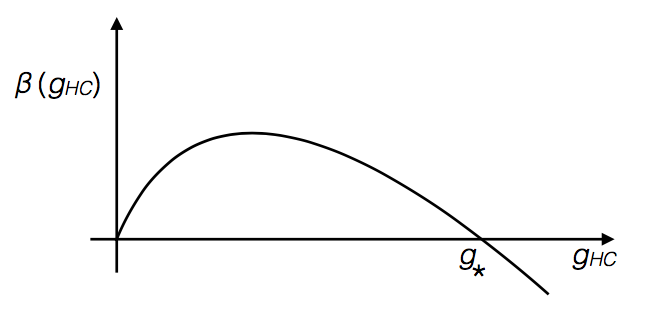}
  \caption{The assumed dependance of the hypercolour $\beta$-function on the gauge coupling. Our model lives in the region $g>g_\star$.}
  \label{beta}
\end{figure}

With the additional matter content, the hypercolour group in our model is no longer asymptotically free as in the original model in Ref.~\cite{Ferretti:2014qta}. As is standard for models employing fermionic partial compositeness~\cite{mchm, wulzer}, we assume instead, that the theory has a strongly coupled  UV  fixed point. This is possible, for instance,  if the $\beta$-function of the hypercolor gauge coupling  has the kind of dependance on the gauge coupling  proposed in Ref.~\cite{holdom} and  shown in Fig.~\ref{beta}. We will assume that our model lives in the region $g>g_\star$ and flows from the UV fixed point $g=g_\star$ to larger values $g> g_\star$ in the IR where it confines.

\subsection{Global symmetry breaking pattern}

When the hypercolor group $SU(4)_{HC}$ confines $\psi$, $\chi^i_{u,d}$ and  $\tilde{\chi}^i_{u,d}$ form TeV-scale condensates, 
\begin{eqnarray}
&&\langle \psi^p \psi^q \rangle \sim \delta^{pq} f_\psi, ~\langle \chi_{u}^{p, i} \tilde{\chi}^{p, j}_{u} \rangle\sim \delta^{ij}f_{\chi_{u}},\nonumber\\&&\langle \chi_{d}^{p, i} \tilde{\chi}^{p, j}_{d} \rangle\sim \delta^{ij}f_{\chi_{d}}
\label{cond}
\end{eqnarray}
 thus breaking the original  global symmetry.  Here $p, q$ are the indices under the $SU(5)$ global symmetry and $i, j$ the indices under the $SU(3)^u_{F}$ or $SU(3)^d_{F}$ flavour symmetry. The condensates break $SU(5)$ to $SO(5)$ and $SU(3)\times SU(3)'$ down to the diagonal $SU(3)_c$ as in Ref.~\cite{Ferretti:2014qta}.
 The coset space, $\GF/\HF$,  is given by,
\begin{eqnarray}\label{coset}
\frac{SU(5) \times SU(3) \times SU(3)' \times SU(3)_{F}^2 \times U(1)^4}{SO(5) \times SU(3)_c \times U(1)_X \times  U(1)_B} \nonumber\\
\end{eqnarray}
where $U(1)^4= U(1)_X \times U(1)_B \times U(1)_{A1}\times  U(1)_{A2}$ and $SU(3)^2_{F}=SU(3)^u_{F}\times SU(3)^d_{F}$.

The unbroken diagonal $SU(3)_c$ is gauged to give the  QCD Lagrangian. The electroweak group is also gauged. It is embedded in $SO(5)$ as follows,
\begin{eqnarray}
&&SO(5) \times SU(3)_c \times U(1)_X   \nonumber\\&&\supset \GCUS \equiv SU(3)_c \times SU(2)_L \times SU(2)_R \times U(1)_X \nonumber\\
&&\supset \GSM \equiv SU(3)_c \times SU(2)_L \times U(1)_Y  \nonumber\\&&\supset SU(3)_c \times U(1)_{\mathrm{e.m.}},
\end{eqnarray}
where  hypercharge generator is   given by $Y= T_{3R}+X$, $T_{3R}$ being the diagonal $SU(2)_R$ generator. 


The Lagrangian in Eq.~\ref{kinetic} is invariant under 5 independent $U(1)$ symmetries, one associated  with the rotation of each of the five fermion species $\chi_u^i, \chi_d^i, \tilde{\chi}_u^i, \tilde{\chi}_u^i$ and $\psi$. Two of these $U(1)$s are not spontaneously broken by the condenstaes in Eq.~\ref{cond}. The first is the linear combinations of these five symmetries that corresponds to  the $U(1)_X$ symmetry in  Table~\ref{fields}.  The second is a linear combination of these    $U(1)$s that can be identified with, $U(1)_B$, the extension of the baryon number symmetry that includes the new  fields. There are three remaining $U(1)$s that get broken spontaneously by the condensates. One linear combination of these three $U(1)$s is anomalous and thus not a true symmetry, which is why it  does not appear in Eq.~\ref{coset}. This still leaves two $U(1)$s,  namely  $U(1)_{A1}$ and $U(1)_{A2}$, shown in Table~\ref{fields}.

We can now count the number of potential Goldstone bosons  that live in the above coset space. There are two  Goldstone bosons, $\eta'_1$ and  $\eta'_2$, corresponding to the spontaneous breaking of $U(1)_{A1}$ and $U(1)_{A2}$, respectively. In the electroweak sector the symmetry breaking pattern,
\begin{eqnarray}
\frac{SU(5)}{SO(5)}
\end{eqnarray}
gives rise to 14 pseudo-Goldstone bosons. These include the Higgs doublet, $H$, a real singlet, a hyprcharge neutral $SU(2)_L$ triplet and a complex $SU(2)_L$ triplet charged under hypercharge.
Finally, in the QCD sector the symmetry breaking pattern,
\begin{eqnarray}
 \frac{SU(3) \times SU(3)'}{SU(3)_c} 
\end{eqnarray}
gives rise to 8 pseudo-Goldstone bosons. As we will discuss shortly, all these Goldstone modes become massive once we introduce other terms in the Lagrangian that explicitly break the original global symmetry, $G_F$.

\begin{table*}[htp]
  \centering
  \begin{tabular}{c|c||c|c|c|c|c|c|c|c|c|}
      & $SU(4)_{HC}$ & $SU(5) $& $SU(3)$ & $SU(3)'$& $SU(3)_F^u$& $SU(3)_F^d$ & $U(1)_X$ & $U(1)_B$ & $U(1)_{A1}$ & $U(1)_{A2}$  \\
      \hline
    $\psi$ & $\mathbf{6}$ & $\mathbf{5}$ & $\mathbf{1}$ & $\mathbf{1}$& $\mathbf{1}$& $\mathbf{1}$ & $0$& $0$ & $-18/5$& 0 \\
    $\chi^i_u $ &$\mathbf{4}$ & $\mathbf{1}$ & $\mathbf{3}$ & $\mathbf{1}$ & $\mathbf{3}$& $\mathbf{1}$& $-1/3$& $-1/6$ & $1$&1\\
    $\tilde\chi^i_u $ & $\bar{\mathbf{4}}$ & $\mathbf{1}$ & $\mathbf{1}$ & $\bar{\mathbf{3}}$ & $\bar{\mathbf{3}}$& $\mathbf{1}$& $1/3$& $1/6$ & $1$&1 \\
    $\chi^j_d $ &$\mathbf{4}$ & $\mathbf{1}$ & $\mathbf{3}$ & $\mathbf{1}$& $\mathbf{1}$& $\mathbf{3}$ & $1/6$ & $-1/6$& $1$&-1\\
    $\tilde\chi^j_d $ & $\bar{\mathbf{4}}$ & $\mathbf{1}$ & $\mathbf{1}$ & $\mathbf{1}$& $\mathbf{1}$ &$\bar{\mathbf{3}}$ & $-1/6$ & $1/6$& $1$ &-1\\
  \hline
  \end{tabular}
  \caption{\small The two-component left-handed  fermions of the UV theory. The confining  hypercolor gauge group is $\GHC$ and $\GF=SU(5)\times SU(3)\times SU(3)'\times SU(3)^u_{F}\times SU(3)^d_{F}\times U(1)_X \times U(1)_B \times U(1)_{A1} \times U(1)_{A2}$ is the global symmetry group before symmetry breaking. }\label{fields}
\end{table*}

\subsection{Partial Compositeness}

We will now discuss how partial compositeness can be realised in this model. We need a composite fermionic partner for each of the SM fermions. In the UV near the fixed point $g=g_\star$ in Fig.~\ref{beta} we identify  the  following baryonic operators, 
\begin{eqnarray}
&&{{\cal O}}^{c,i}_{uL}=({\chi}_u\psi{{\chi}_u})^i~~~~
{{\cal O}}^{i}_{uR}=(\bar{{\chi}}_u\bar{\psi}{\tilde{\chi}}_u)^i\nonumber\\
&&{{\cal O}}^{c,i}_{dL}=({\chi}_d\psi{{\chi}_d})^i~~~~
{{\cal O}}^{i}_{dR}=(\bar{{\chi}}_d\bar{\psi}{\tilde{\chi}}_d)^i,
\label{baryonComb}
\end{eqnarray}
which have components that have the right transformation properties to be partners of right-handed up type quarks, left-handed up type quarks, right-handed down type quarks and left-handed down type quarks respectively  These are all left-handed two component spinor states that transform as triplets under  $SU(3)_c$ and the $SU(3)^{u,d}_{F}$ flavour group.

The operator ${{\cal O}}^i_{uR}$ transforms as $(\mathbf{5}, \mathbf{3})_{2/3}$ under $SO(5) \times SU(3)_c \times U(1)_X$.  It has components, that we call $U^i_R$,  which transform as $(\mathbf{3},\mathbf{1})_{2/3}$ under $SU(3)_c \times SU(2)_L \times U(1)_Y$ that can be identified as the partner for $u_R^{c,i}$, the   antiparticle for the SM  right-handed up quarks. Similarly ${{\cal O}}^{c, i}_{uL}$ transforms as $(\mathbf{5},{\bar{ \mathbf{3}}})_{-2/3}$ under $SO(5) \times SU(3)_c \times U(1)_X$ and  has components that transform as $(\bar{\mathbf{3}},\mathbf{2})_{-2/3}$ under $SU(3)_c \times SU(2)_L \times U(1)_Y$. We will call these components,  $U^{c, i}_{L}$, and they will serve as partners for the SM left handed up-type fermions. 

As far as, ${{\cal O}}^j_{dR}$ and
${{\cal O}}^{c, j}_{dL}$, are concerned they transform respectively as $(\mathbf{5}, \mathbf{3})_{-1/3}$ and  $(\mathbf{5}, \bar{\mathbf{3}})_{1/3}$  under  $SO(5) \times SU(3)_c \times U(1)_X$. They, have components that transform respectively as  $(\mathbf{3},\mathbf{1})_{-1/3}$  and  $(\bar{\mathbf{3}},\mathbf{2})_{1/3}$ under $SU(3)_c \times SU(2)_L \times U(1)_Y$; we will call these $D^j_R$ and $D^{c,j}_L$, the partners for the right-handed anti-down quarks and the left handed down quarks.

The partial compositeness Lagrangian can now be realised by linearly  coupling the SM fermions to their partners,
\begin{eqnarray}\label{uvlag}
{\cal L}_{mix}&=&\frac{\lambda^{ij}_{u_R}}{4 \pi}\frac{1}{\Lambda^{d_{U_R}-5/2}} u_R^{c,i} U^j_R+\frac{\lambda^{ij}_{u_L}}{4 \pi}\frac{1}{\Lambda^{d_{U_L}-5/2}}q^i_L U^{c,j}_L \nonumber\\
&+& 
\frac{\lambda^{ij}_{d_R}}{4 \pi}\frac{1}{\Lambda^{d_{D_R}-5/2}}d_R^{c,i}D_R^j +\frac{\lambda^{ij}_{d_L}}{4 \pi} \frac{1}{\Lambda^{d_{D_L}-5/2}} q^i_L D^{c,j}_L\nonumber\\&+& h.c.
 \end{eqnarray}
where $d_F$ is the conformal dimension of the corresponding  operator, $F$. The conformal dimensions are independent of the flavour indices $i, j$ because of the $SU(3)^u_{F}\times SU(3)^d_{F}$ symmetry.  The coupling of the SM fermions to the other possible baryonic operators-- such as the right-handed SM quarks with $(\bar{\tilde{\chi}}_{u,d}\psi\bar{\tilde{\chi}}_{u,d})^i$ and $({{\chi}}_{u,d}\psi{{\chi}}_{u,d})^i$ or  the left-handed quarks with $(\bar{\tilde{\chi}}_{u,d}\psi\bar{\tilde{\chi}}_{u,d})^i$ and $(\bar{\tilde{\chi}}_{u,d}\bar{\psi}{\chi}_{u,d})^i$ --  is prohibited  as we impose a $Z_2$ symmetry under which ${\chi^i_{u,d}}$ and the SM right-handed fermions are odd.

In the IR after confinement the above operators lead to  composite states that pair with  charge conjugates states (that can be  obtained by interchanging $\chi_{u,d}\leftrightarrow\tilde{\chi}_{u,d}$) to form massive Dirac fermions. Once these are integrated out the SM Yukawa coupling between the SM fermions and the composite Higgs boson is generated (see for eg. Ref. \cite{wulzer}).
\begin{figure*}[htp]
  \centering
  \subfigure[]{\includegraphics[scale=0.4]{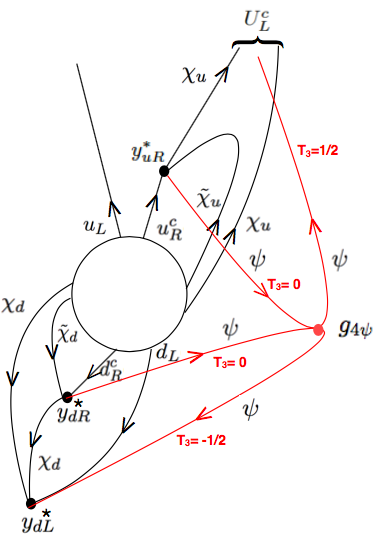}}\quad
  \subfigure[]{\includegraphics[scale=0.4]{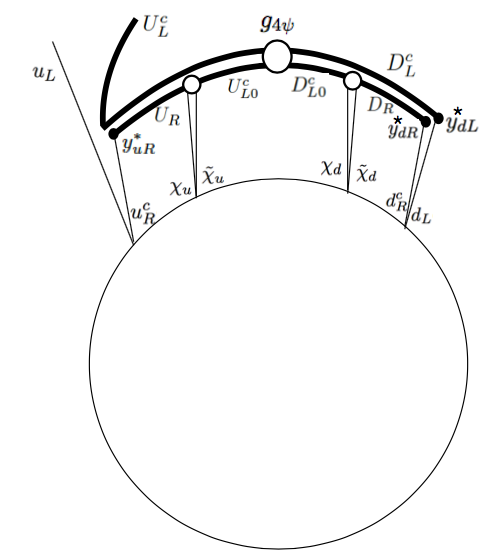}}
  \caption{(a) Diagram showing the QCD instanton contribution  to generate $u_L U^c_{L}$ including only the first generation of fermions; the generation indices for the hypercolour fermions have been omitted for convenience. An identical topology exists for the other generations with the only difference that the $c_L$-$U_L^{c2}$ and $t_L$-$U_L^{c3}$ lines are closed by the couplings $y^*_{cL}$ and $y^*_{tL}$. The red lines do not intersect with the others. (b) The same diagram drawn in terms of lines of hypercolour and QCD singlets. 
The white circles represent unsuppressed vertices arising from the strong sector whereas the dark circles denote vertices  suppressed by couplings external to the strong sector.}
\label{insta}
\end{figure*}
\subsection{Explicit breaking by masses and four-$\psi$ interaction }
Finally we add some additional terms not present in the original model of Ref.~\cite{Ferretti:2014qta},
\bea
{\cal L}_{new}&=&m_{0}e^{i \theta_m}\psi_0 \psi_0 +\frac{g_{4\psi}e^{i \theta_g}}{\Lambda^2} (\bar{\psi}_{-+}\bar{\sigma}_\mu \psi_0)(\bar{\psi}_{+-} \bar{\sigma}^\mu \psi_0)\nonumber\\
\label{new}
\eea
where   $m_0$ and $g_{4\psi}$ are real,    and the subscript $\{\alpha,\beta\} $ in $\psi_{\alpha \beta}$ in the last term refers to the $T_{3R,3L}$ charges. To  make our notation clear and to understand how each of these components of $\psi$ transform, recall that fermion $\psi$ transforms as a \textbf{5} of $SU(5)$. This decomposes into two $SU(2)_L$ doublets and an $SU(2)_L$ singlet,
\begin{equation}\label{psivec}
\left(\begin{array}{c}
	\Psi_+ \\
	\Psi_- \\
	\psi_0
\end{array}\right)\,,
\end{equation}
$\Psi_{\pm}$  are SU(2)$_L$ doublets and the $\pm$ subscripts correspond to $T_{3R}=\pm1/2$.  We can explicitly write $\Psi^{T}_{\pm}=\left(\psi_{\pm +}, \psi_{\pm -}\right)$ where the second index now corresponds to the $T_{3L}=\pm1/2$. 

The first term breaks $U(1)_{A1}$ and thus   gives a mass to $\eta'_1$.    The  $\eta'_1$ does not get a contribution form the mixing terms  in Eq.~\ref{uvlag} as  we can extend the $U(1)_{A1}$ symmetry to the SM fermions  in a way that is preserved by  Eq.~\ref{uvlag}, i.e. by giving the $U(1)_{A1}$ charges, $8/5$ and $18/5$, respectively to the SM doublet and singlet fermions.   The  $\eta'_1$ actually  would  eventually get a contribution to its mass also from non-perturbative QCD effects as the above extended $U(1)_{A1}$  is anomalous under QCD. On the other hand, $U(1)_{A2}$ is already broken by the mixing terms in Eq.~\ref{uvlag} which give $\eta'_2$ a mass.

 The second term in Eq.~\ref{new} is a four-$\psi$ interaction between components of the fermion $\psi$ that explicitly breaks the original global symmetry $SU(5)$. This term would be essential in enhancing the QCD small instanton contributions in the next section.


\subsection{A minimally flavour and CP violating strong sector}
\label{cp}

Notice that the lagrangian in Eq.~\ref{kinetic} is invariant under CP and the $SU(3)^u_{F}\times SU(3)^d_{F}$  flavour symmetriy.  These symmetries are broken only by the couplings $\lambda^{ij}_{uL}, \lambda^{ij}_{uR}, \lambda^{ij}_{dL}, \lambda^{ij}_{dR}, g_{4 \psi} e^{i \theta_g}, m_\phi e^{i \theta_m}$ and  the strong CP phases in the QCD and hypercolor sectors, $\theta_{QCD}$ and $\theta'$ respectively. Following, Ref.~\cite{redi} here we will further assume that the mixings of the right handed quarks do not break the $SU(3)^u_{F}\times SU(3)^d_{F}$  symmetry so that, 
\beq
\lambda^{ij}_{uR, dR} \sim  y_{uR, dR}~e^{i \theta_R}\delta^{ij}.
\label{RH}
\eeq
This implies that the SM  Yukawa couplings would be proportional to the left handed mixings,
\bea
&&Y^{ij}_{u} \sim \frac{\lambda^{ik}_{uL} \lambda^{kj}_{uR}}{4 \pi }\sim  \lambda^{ij}_{uL} y_{uR}\nonumber\\
&&Y^{ij}_{d} \sim\frac{\lambda^{ik}_{dL} \lambda^{kj}_{dR}}{ 4 \pi}\sim \lambda^{ij}_{dL} y_{dR}
\label{yukawa}
\eea
which are the only spurions that break flavour symmetry, thus realising minimal flavour violation (MFV)~\cite{DAmbrosio:2002vsn}.

As far as CP phases, $\theta_m, \theta_g, \theta_R, \theta_{QCD}$ and $\theta'$   are concerned, we can transfer all of them to $\lambda^{ij}_{uL}$ and $\lambda^{ij}_{dL}$ by taking the following steps:
\begin{enumerate}
\item{First, the phase $\theta_m$ can be rotated away by  ${\psi}_{0}\to {\psi}_{0} e^{-i \theta_m/2}$ which redefines $\theta_g, \theta_R$ and $\theta'$.}
\item{ Next, the phase   $\theta_g$ associated to $g_{4 \psi}$  can be rotated to $\lambda^{ij}_{uL}$ and $\lambda^{ij}_{dL}$ by making the transformation ${\psi}_{-+}\to {\psi}_{-+} e^{i \theta_g}$. This also redefines $\theta'$.}
\item{ Then  $\theta'$ can be eliminated by an equal rotation of all $\chi_i$ and $\tilde{\chi}_i$, which also redefines $\theta_{QCD}$.}
\item{Finally   $\theta_R$ can be eliminated by an equal but opposite  rotation of  the $\chi_i$ relative to the  $\tilde{\chi}_i$.}
\end{enumerate}
 This still leaves $\theta_{QCD}$  which can be entirely shifted to  $\lambda^{ij}_{uL}$ and $\lambda^{ij}_{dL}$ by chiral rotations of the SM quarks while keeping the combination, 
 \bea
 \bar{\theta}_{QCD}=\theta_{QCD}+ {\rm Arg}{\rm Det}\[\lambda_u \lambda_d\]
 \eea
 unchanged. Because we have a MFV like structure, as in the SM, there is only one more physical phase in our theory, the CKM phase, 
\bea
 {\theta}_{CKM}= {\rm Arg}{\rm Det}\[\lambda_u \lambda_d-\lambda_d\lambda_u\].
 \eea


\section{Effect of small instantons}
The effect of QCD  instantons at high energies are suppressed due to two reasons, (1) the  suppression factor $\kappa_s=e^{-2 \pi/\alpha_s}$ is small as QCD is asymptotically free, and,  (2) there is a suppression factor that goes as the product of the Yukawa couplings of all the SM quarks, all of which are active at high energies. Both these effects can be overcome in the model we are considering because, (1) the new coloured fermions $\chi^i_{u,d}$ and  $\tilde{\chi}^i_{u,d}$ that  form the composite fermionic partners can lead to large UV values of the QCD coupling  and (2) the mixings, $ \lambda^{ij}_{uL, dL}$, and thus the Yukawas in Eq.~\ref{yukawa}   can run to higher  values in the UV in these models as  we will now show.
\subsection{Running   of $\alpha_s$}
In our model there are  8 new flavours of fermions for every generation once we take into account the 4 hypercolour degrees of freedom of $\chi_i^{u,d}, \tilde{\chi}_i^{u,d}$. Including the SM fermions, there are $n_f=30$ flavours. Using the usual expression,
\beq
\frac{d g_s }{d \log \mu}= -(11-2 n_f/3)\frac{g_s^3}{16 \pi^2}
\eeq
we find that the QCD beta function is positive. Assuming that the new flavours become active at 1 TeV  we find that $g_s= 4 \pi$ for $\mu \sim 2000$ TeV  where the instanton vertex will become unsuppressed.

We will assume that some UV degrees of freedom cut-off this growth of this coupling above a scale $M\sim 2000$ TeV such that  $\kappa_s$ has a maximal value at this scale. We will treat this maximal value as a free parameter that can vary from $\kappa_s=10^{-34}$ for $g_s=1$ to  $\kappa_s\sim 1$ for $g_s= 4 \pi$. We will also assume that at a scale $M'>M$, the QCD gauge coupling growth is tamed, $\kappa_s$ becomes negligible and the QCD instantons are again highly suppressed.

\subsection{Running of $\lambda^{ij}_f$}
 We will work in the mass basis where $\lambda^{ij}_{uL}(M')={\rm diag}\[y_{uL}, y_{cL},y_{tL}\]$ and $\lambda^{ij}_{dL}(M')={\rm diag}\[y_{dL}, y_{sL},y_{bL}\]$. The couplings  $y_f$ run between the UV and IR scale,  $m_\star \sim$ 1 TeV,  of the composite masses, 
\beq
\mu\frac{d y_{f}}{d  \mu} =(d_F-5/2)y_f+b  \frac{N_{HC} \, y_{f}^3}{16 \pi^2}
\label{rgyukawa}
\eeq
where $b$ is an ${\cal O}(1)$ factor, $N_{HC}=4$ is the number of colours for the hypercolour group and  $d_F$ is the conformal dimension of the operator corresponding to the fermionic partner, $F$,  that couples to the SM fermion, $f$. The first term allows an anarchic and ${\cal O}(1)$ valued matrix $\lambda^{ij}_{f}(M)$ to generate a hierarchical $\lambda^{ij}_{f}(m_\star)$ thus explaining the SM masses and mixings. This can be seen if we  solve the above equation by ignoring the second term,
\beq
y_{f}(m_\star)=y_{f}(M)\left(\frac{m_\star}{M}\right)^{d_F-5/2},
\label{run}
\eeq
which shows that ${\cal O}(1)$ differences in the $d_j$ can lead to exponential hierarchies in the IR. Including the second term does not change this qualitative feature, in fact it can lead to hierarchies between couplings involving operators with the same $d_F$. In particular it results in a fixed point at $y_f =4 \pi/\sqrt{b N_{HC}  \gamma_F}$ where $\gamma_F= d_F- 5/2$.




\section{Up quark mass from small instantons}
\label{sec:upmass}

 In this section we will consider the model defined in Sec.~\ref{model} and assume  that one of the eigenvalues of $\lambda^{ij}_{uL}$  vanishes at the scale $M'$  where  instanton effects are negligible, i.e.,  $y_{uL}(M')=0$.  Instanton effects around the scale $M$  then generate a non-zero value for the up quark mixing, $y_{uL}$, via the 't Hooft vertex.

\subsection{Non-perturbative generation of $y_{uL}$} 

The 't Hooft instanton vertex due to the QCD anomaly in this model is an interaction including all the coloured fermion species. In Fig.~\ref{insta} (a) we show only the first generation fermions, $u_L, u^c_R,d_L, d^c_R,{\chi}^1_u,\tilde{\chi}^1_u,{\chi}^1_d$ and $\tilde{\chi}^1_d$ explicitly. Working in the mass basis we start  from the anomaly vertex to  generate the $y_{uL} u_L U^{c1}_L$  term as shown in  Fig.~\ref{insta} (a).  The   fermions of the other generations  have not been shown for space constraints but a identical topology exists for them  with the only difference  that now the $c_L$-$U_L^{c2}$ and $t_L$-$U_L^{c3}$ lines are also closed by the couplings $y^*_{cL}$ and $y^*_{tL}$. 

 To get the  NDA estimate for $y_{uL}$ we can redraw the same diagram but now in terms of QCD and hypercolour singlets as shown in  Fig.~\ref{insta} (b). If one considers the pairs, $U_L^{c1} U_R^1$, $U_L^{c1} U_{L0}^{c1}$, $D_L^{c1} D^1_R$ and $D_L^{c1} D_{L0}^{c1}$ as QCD singlet scalars, this diagram becomes very similar to the one considered in Ref.~\cite{randall} where new scalars connect fermion pairs, such as $u_L u_R^c$ to $d_L d_R^c$, in the 't Hooft vertex. The result of a full calculation in Ref.~\cite{randall} is that the only suppression factor is given by,  $\prod_f {{\cal Y}_f}/{4 \pi}$, where the ${\cal Y}_f$ are the Yukawa couplings of the scalars to the fermion pairs. In our case the coupling of the SM fermions to the scalars, formed from the composite partners, reaches its perturbative limit for $y^{*}_{fL}= y^*_{fR}= 4 \pi$. Thus adapting the result  of Ref.~\cite{randall} to our case, and including a suppression factor corresponding to $g_{4\psi}$, we obtain, 
\bea
\frac{y_{uL}}{4 \pi} &\sim& \kappa_s\left(\frac{g_{4\psi}}{16 \pi^2}\right)^3 \frac{y^{*}_{uR}}{4 \pi}\prod_{f=d,s,c,b,t}\frac{y^{*}_{f_L}}{4 \pi}\frac{y^*_{f_R}}{4 \pi},
\label{delyur}
\eea
where all the above couplings are at the scale $M$, and  following Eq.~\ref{RH}, all the $y_{fR}= y_{uR,dR}$ depending on whether $f$ is an up or down type fermion. The white circles in  Fig.~\ref{insta}(b) represent vertices arising from the strong sector whereas the dark circles denote vertices external to the strong sector. Here we have assumed a suppression only due to the former couplings

The known value of the up quark Yukawa can be reproduced  in the strongly coupled regime  when the couplings in Eq.~\ref{delyur} saturate their perturbative limit. For instance we obtain for the up Yukawa, 
\bea
Y_u (m_\star)\sim\frac{y_{uL} (m_\star)y_{uR}(m_\star)}{4 \pi}&\sim& 1.5 \times 10^{-5} \kappa_s\left(\frac{g_{4\psi}}{16 \pi^2}\right)^3\nonumber\\
\label{delyurnum}
\eea
if we solve the RG equations in Eq.~\ref{rgyukawa} assuming $b=1/4$,  taking $d_{U_L, D_L}=7/2$,   $y_{uR, dR}(M)=4 \pi$ in Eq.~\ref{RH}, $y_{uR,dR}(m_\star)=y_{uR, dR}(M)/10$, and other boundary conditions  fixed by the measured value of the SM fermions masses.  



%

\subsection{Solution to the strong CP problem}
  Let us first consider the scale $M'$ at and above which QCD instanton effects are suppressed so that  $y_{uL}(M')=0$. At this scale the phase  $\theta_{QCD}$ can be removed simply by a chiral rotation  of the up quark fields.
  
  It is still instructive to see how, at leading order,  $\theta_{QCD}$ vanishes   just below the scale $M$  where  QCD instanton effects become important. These effects generate a non-zero $y_{uL}$ given by Eq.~\ref{delyur}.  As explained in Sec.~\ref{cp}, we are working in a convention where chiral rotations are used to transfer all the phases   to the mixing matrices $\lambda_{f}^{ij}$ and there is no $G \tilde{G}$ coupling to start with.  A  contribution to $\bar{\theta}_{QCD}$  from the fermionic phases can arise at this scale from closing the 't Hooft vertex completely, which gives, 
\bea
\bar{\theta}_{QCD}(M)&=&{\rm Arg}(y^*_{uL}y^*_{uR} \prod_{f=d,s,c,b,t}y^{*}_{fL}y^*_{fR})\nonumber\\
&=&{\rm Arg}(\prod_{f=u,d,s,c,b,t}|y_{fL}|^2|y_{fR}|^2)=0.
\eea

As explained in Sec.~\ref{cp} there is only one more physical phase in our theory, the CKM phase,  ${\theta}_{CKM}$. From the scale $M$ to the experimental scale,  ${\theta}_{CKM}$ can induce $\bar{\theta}_{QCD}$ due to Renormalisation Group (RG) effects, but as in the SM this is expected to be highly suppressed. This is because the arguments based on spurion analysis  for the SM in Ref.~\cite{Ellisgaillard} can be adapted to our model to show that this effect is at least 7-loop suppressed \footnote{The diagrams shown  in Ref.~\cite{Ellisgaillard}  for the SM would apply for our model as long as the  Higgs fermion couplings in the SM diagram are appropriately  replaced by strong dynamics. Apart from RG effects there are also finite contributions to the strong CP phase from the CKM phase, but in Ref.~\cite{Ellisgaillard} these were also shown to be much smaller than the experimental bound, $\bar{\theta}_{QCD}\lesssim 10^{-10}$, from neutral electric dipole moment experiments~\cite{Pospelov, Baker}.}

 \section{A heavy axion  from small instantons}
 \label{axion}

Now we consider a different  scenario from Sec.~\ref{sec:upmass}, taking $y_{uL}(M') \neq 0$, but introduce a new pseudoscalar field, $\phi$. In addition to  all the Lagrangian  terms in Sec.~\ref{model}, we consider the new term, 
\bea
{\cal L}_\phi=\frac{g_s^2}{32 \pi^2}\frac{\phi}{f}G_{\mu \nu}\tilde{G}^{\mu \nu}.
\eea
where we take  $f\gtrsim M \sim 10^{6}$ GeV. We assume that  above term is the only one that breaks the shift symmetry of $\phi$ via anomalous QCD effects so that it can be identified with the QCD axion.

 Non-perturbative effects at the scale $M$ in the UV as well as the usual QCD scale in the IR  will contribute to the axion potential. The UV contribution  can be estimated by first shifting $\bar{\theta}_{QCD}$ to the 't Hooft vertex in  Fig.~\ref{insta} (a) and \ref{insta}(b) and then  completely  closing all the fermion lines  which is possible now that  $y_{uL} \neq 0$.  This then gives a new contribution to the axion potential,
\bea
V(\phi)= \kappa M^4 \cos \left(\frac{\phi}{f}+\bar{\theta}_{QCD}\right)+ m_\pi^2 f_\pi^2 \cos \left(\frac{\phi}{f}+\bar{\theta}_{QCD}\right)\nonumber
\eea
where, 
\bea
\kappa\sim \kappa_s \left(\frac{g_{4\psi}}{16 \pi^2}\right)^3 \prod_{f=u,d,s,c,b,t}\frac{y^{*}_{fL}}{4 \pi}\frac{y^*_{fR}}{4 \pi}.
\eea
Again, all the above couplings are at the scale $M$ and all the $y_{fR}= y_{uR,dR}$ depending on whether $f$ is an up or down type fermion (see Eq.~\ref{RH}). The second term above is the usual large instantion contribution in the IR, with $m_\pi$ being the pion mass and $f_\pi$, the pion decay constant. Both the terms are aligned in phase because both the contributions arise from closing the same 't Hooft vertex, but at different scales.

\begin{figure}[t]
  \centering
  \includegraphics[scale=0.5]{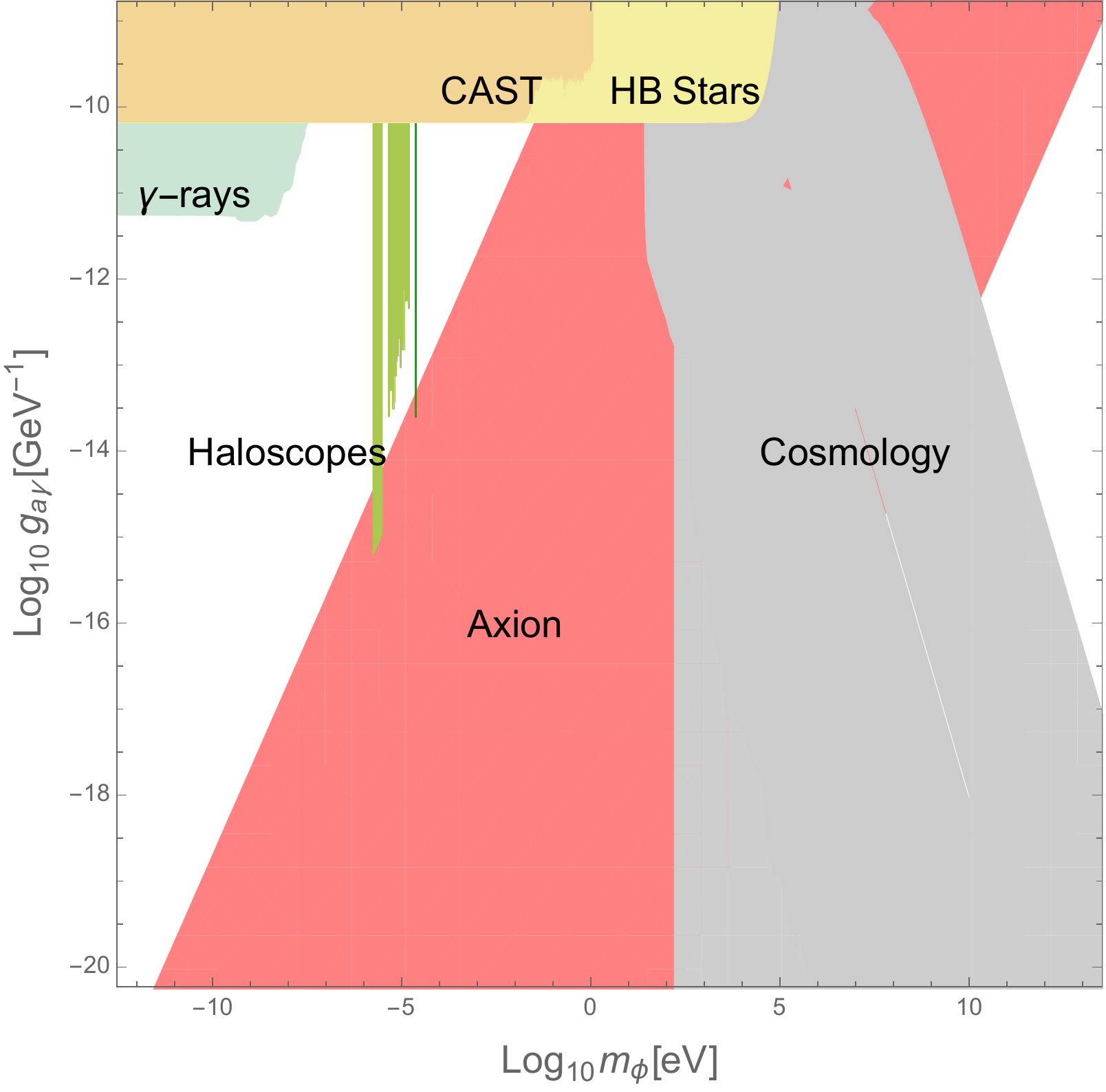}
  \caption{The allowed parameter space for the QCD axion of Sec.~\ref{axion} is shown in red. The other  bounds have been adapted from Ref.~\cite{redondo, agrawal1}  where a detailed discussion of these can be found.}
  \label{alps}
\end{figure}

Solving the RG equations in Eq.~\ref{rgyukawa} assuming $b=1/4$,  taking $d_{U_L, D_L}=7/2$,  $y_{uR,dR}(M)=4 \pi$ in Eq.~\ref{RH}, $y_{uR,dR}(m_\star)=y_{uR, dR}(M)/10$, and other boundary conditions  fixed by the measured value of the SM fermions masses, we obtain, 
\bea
\kappa\sim \kappa_s \left(\frac{g_{4\psi}}{16 \pi^2}\right)^3 4.2 \times 10^{-4}.
\label{maximal}
\eea
The above numerical value is the maximal possible one corresponding to the case  when all the couplings saturate their perturbative limit. The factor, $\kappa_s$, can  vary over a large range from unity to exponentially small values as the strong coupling is varied; the axion mass can thus vary from its minimum value due to the IR contribution to  values as large as 10 TeV when the suppression factor, $\kappa$ approaches the maximal value in Eq.~\ref{maximal}.

 We show the allowed region in the coupling-mass parameter space in Fig.~\ref{alps}, where the $y$-axis is $g_{a\gamma}$ the axion coupling to photons, defined by the coupling, 
\bea
-\frac{g_{a\gamma}}{4} \phi F_{\mu \nu} \tilde{F}^{\mu \nu}.
\eea
where,
\bea
g_{a\gamma}\sim \frac{\alpha_{em}}{2 \pi f }.
\eea
We see that a huge area in the parameter space is allowed for the QCD axion in this model. The usual QCD axion band is  the left edge of the area shown in Fig.~\ref{alps}. While parts of this area are ruled out by existing constraints-- in particular cosmological ones in the region where a thermal population of the axion can exist~\cite{redondo}-- large parts of the allowed region still remain unconstrained.

 \section{Conclusions}
 
 In this work we showed that small instanton effects can become very important in  standard UV completions of composite Higgs models with partially composite fermions. This is possible because both the QCD gauge coupling and effective Yukawa interactions run to larger values with energy resulting in unsuppressed instanston contributions in the UV. As far as the Yukawa interactions are concerned,  it is well-known that  in partially composite models, the hierarchical nature of the SM fermion masses and mixings can arise from anarchic and ${\cal O} (1)$ interactions in the UV. The QCD coupling grows because  in order to have fermionic partners for all SM fermions, many new fermions in the hypercolour sector need to be introduced. These fermions are also charged under QCD and this generates a positive $\beta$-function which results in the QCD gauge coupling running to non-perturbative values in the UV.  The only modification of the lagrangian required to achieve this effect are the explicit breaking  terms  in Eq.~\ref{new}.
 
 As a consequence of this enhancement of small instanton contributions, we show that  the up quark mass can arise entirely from instanton contributions. This implies that the up quark mass vanishes in the deep UV and is only generated additively by instanton effects. In the deep UV one can thus rotate away the strong CP phase.  In an alternative scenario where an axion field exists, we show that its mass   can be as large as 10 TeV.  The allowed parameter space is much larger than the usual QCD band as shown in Fig.~\ref{alps}.  Our model thus  opens up new areas in the coupling-mass parameter space that are still unconstrained by existing bounds. This motivates the development of new experimental strategies to probe these regions.
\bigskip

\noindent
\textbf{Note added:} While we were in the process of completing this project, Ref.~\cite{gher} appeared, that also provides a heavy axion candidate in composite models. While this work also utlises the set-up of Ref.~\cite{Ferretti:2014qta}, there is no overlap  with Sec.~\ref{axion}. This is because  Ref.~\cite{gher} does not use the same mechanisms for enhancing small instantion contributions, namely running of the QCD gauge coupling and the $\lambda^{ij}_f$, that have been used in this work. The other important differences include the fact that Ref.~\cite{gher} has an in-built axion candidate and, unlike this work,  it requires a hypercolour condensation scale of at least  1000 TeV resulting in  a tuned Higgs sector. 
%
%
%
%
%
\section*{Acknowledgements} We would like to thank A. Pomarol for his feedback on this work and for some key suggestions on the model.

\bibliographystyle{utphys}
\bibliography{references} 
\end{document}